\documentclass[a4paper,10pt]{article}
\usepackage{cmap}
\usepackage[T1,T2A]{fontenc}
\usepackage{amsmath,amsthm,amssymb}
\usepackage{mathtext}
\usepackage[utf8]{inputenc}
\usepackage[english,main=russian]{babel}
\usepackage{indentfirst}
\usepackage{setspace}
\usepackage[all]{xy}
\usepackage{paralist}
\usepackage{lineno}
\usepackage{hyperref}
\allowdisplaybreaks[1]
\usepackage{graphicx}
\usepackage{cite}
\usepackage{mathrsfs}
\usepackage[textwidth=14cm,textheight=20cm]{geometry}
\usepackage{booktabs}
\usepackage{comment}
\begin{document}
\noindent
УДК 517.928.4:[519.872.2+519.862.4]\\
MSC 90B10, 90B22, 90B30, 92B05

\begin{center}
{\large Клиринговая функция в контексте метода инвариантного многообразия}%
\end{center}
\textit{А.\,Т. Мустафин}$^1$, \textit{А.\,К. Кантарбаева}$^2$\\
\small $^1$ Казахский национальный исследовательский технический университет им.~К.\,И.~Сатпаева, Казахстан, 050013, г.~Алматы, ул.~Сатпаева, 22\\
\small $^2$ Казахский национальный университет им.~аль-Фараби, Казахстан, 050040, г.~Алматы, пр.~аль-Фараби, 71
\bigskip

Клиринговая функция (КФ) выражает зависимость выпуска от объема производственного ресурса, находящегося в работе (WIP). Применение КФ представляется многообещающим при моделировании для целей планирования такого показателя, как время производственного цикла. В теории очередей аналитическое выражение для КФ выводится в предположении о стационарности процесса производства. Однако в связи с конечной продолжительностью планового периода возникает вопрос о корректности использования понятия КФ в неустановившемся режиме. В настоящей работе применяется альтернативный подход для построения механистической модели производственного цеха, занятого выпуском одного продукта из одного ресурса, основанной на аналогии между работой машины и биологического фермента. Модель приводится к сингулярно возмущенной системе двух обыкновенных дифференциальных уравнений для медленной (WIP) и быстрой (популяция занятых машин) переменных. Анализ этой разнотемповой системы показывает, что КФ есть не что иное как результат асимптотического разложения медленного инвариантного многообразия. Корректность использования КФ в конечном счете определяется малостью параметра, стоящего перед производной от быстрой переменной. Показано, что достаточно малое соотношение <<машины\,:\,WIP>> практически гарантирует применимость приближения КФ в нестационарном режиме работы производства.

\textit{Ключевые слова:} незавершенное производство, модель производства, квазистационарное приближение, сингулярные возмущения, ферментативный катализ

\bigskip
\normalsize
\textbf{1. Введение}

Клиринговая функция (clearing function, далее --- КФ) --- относительно новый термин в исследовании операций и операционном менеджменте. Не следует смешивать его с исторически более ранним финансовым термином <<клиринг>>, который означает систему безналичных расчетов, основанную на зачете взаимных требований и обязательств.

Мотивация к введению понятия КФ, формальное определение которой будет дано ниже, связана с практическими потребностями планирования и прогнозирования производства. В частности --- с необходимостью получения более надежного ответа на наиболее часто задаваемый вопрос потребителей товара или услуги: <<Когда будет готов заказ?>> \cite{Vacanti_2020}.

Теория массового обслуживания, аппарат которой используется для моделирования производства, оперирует с тремя основными показателями очереди: работой, находящейся в процессе выполнения (work in progress), временем цикла (cycle time) и выпуском (throughput). Следуя традиции операционного менеджмента, эти показатели мы будем обозначать соответствующими аббревиатурами WIP, CT и TH.

Работа в процессе выполнения, которую часто также называют <<незавершенным производством>>, означает объем материалов и заготовок, вовлеченных в производство, превращение которых в продукт пока не начиналось или еще не закончено. С одной стороны, WIP --- лучший предиктор общей эффективности производственной системы. С другой стороны, две прочих характеристики, CT и TH, так или иначе определяются в терминах WIP.

Время цикла --- это среднее время от постановки работы в очередь до превращения в продукт или, иначе говоря, время, которое работа находится в статусе WIP.

Наконец, выпуск определяется как объем продукции данного типа, произведенной за единицу времени, то есть количество работы, законченной за условный единичный период.

В динамически равновесном режиме производства входящий поток заданий (работ) равен исходящему потоку продукции. Согласно закону Дж.~Литтла \cite{Little_1961, Little_2008}, средние величины WIP, CT и TH связаны между собой соотношением CT = WIP/TH. Помимо равенства средних потоков на входе и на выходе также предполагается, что а)~каждое поступившее задание когда-нибудь будет выполнено и покинет систему, не потерявшись; б)~объем незавершенной работы примерно одинаков в начале и конце рассматриваемого отрезка времени; в)~средний возраст незавершенной работы не увеличивается и не уменьшается. Формула Литтла замечательна своей общностью, поскольку не предполагает стационарности случайных процессов поступления заданий и их обработки и не требует дополнительных сведений о вероятностных характеристиках потоков, о количестве машин, участвующих в производстве, или о порядке выполнения работ.

Подобно термодинамическому тождеству закон Литтла позволяет однозначно найти по любым двум заданным показателям состояния системы третий. К сожалению, с той же степенью общности никак нельзя связать рассматриваемые величины попарно. Неизбежно приходится либо выводить специальную модель, отталкиваясь от характеристик изучаемой очереди, либо строить нужную регрессионную модель по эмпирическим данным.

КФ определяется как детерминистическое соотношение между выпуском продукции за некоторый период времени и работой, находящейся в процессе выполнения в тот же период: TH(WIP). КФ как раз и предназначена для прогнозирования времени цикла по заданному объему незавершенной работы. Такое название функция получила, потому что характеризует способность системы <<очищаться>> (clear) от незавершенной работы.

По существу, КФ представляет собой особую разновидность производственной функции в представлении <<запас--поток>> в отличие от обычной формы <<поток--поток>>, используемой в методе межотраслевого баланса. Аппарат КФ возник в рамках <<жидкостного>> (гидродинамического) приближения в теории массового обслуживания (напр., обзор \cite{Armbruster_2012a}). В этом подходе случайный поток дискретных работ/заданий (например, деталей или иных обрабатываемых предметов) заменяется --- в пределе закона больших чисел --- непрерывным и детерминированным потоком бесконечно делимых частиц, подобным потоку жидкости.

Как таковая, хотя и не под современным названием, КФ была впервые введена С.~Грейвсом \cite{Graves_1986}, который предположил прямую пропорциональную зависимость выпуска от WIP. Несмотря на удобство для вычислений данный подход оказался не в состоянии описать наблюдаемый на практике эффект насыщения выпуска. Минимальная модель, удовлетворяющая требованиям монотонности, вогнутости и насыщения, вместе с самим термином КФ была предложена У.~Кармаркаром \cite{Karmarkar_1989, Karmarkar_1993} в виде гиперболической зависимости
\begin{equation}\label{E:Karmarkar}
\mathrm{TH} = \frac{\widehat{\mathrm{TH}}\ \mathrm{WIP}}{K +\mathrm{WIP}},
\end{equation}
где $\widehat{\mathrm{TH}}$ --- максимально возможный выпуск, $K$ --- константа полунасыщения, такая, что $\mathrm{TH} = \frac{1}{2}\widehat{\mathrm{TH}}$ при $\mathrm{WIP}=K$. Формула \eqref{E:Karmarkar} выведена для производственной очереди $M/M/1$ с пуассоновским входным потоком, экспоненциально распределенным временем обработки и одной обслуживающей машиной. А.~Сринивасан \cite{Srinivasan_1988} рекомендовал эмпирическую экспоненциальную зависимость
\begin{equation*}
\mathrm{TH} =\widehat{\mathrm{TH}}\left(1-\mathrm{e}^{-a\,\mathrm{WIP}}\right),
\end{equation*}
где $a$ --- подгоночный параметр. Хотя за прошедшие тридцать лет были предложены и другие аналитические формулы КФ разной степени сложности (см. обзор \cite[ch.~7, pp.~143--189]{Missbauer_2020}), формула Кармаркара в настоящее время наиболее популярна.

\begin{figure}[htb]
\noindent\centering{%
\includegraphics[scale=1]{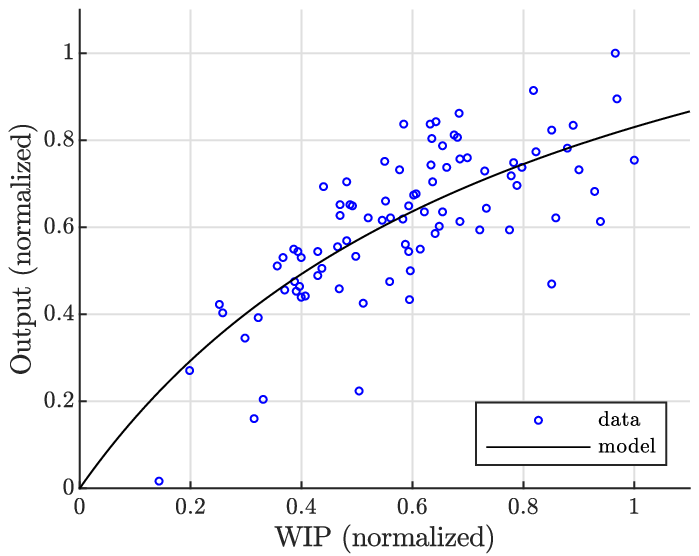}}
\caption{Сравнение теоретической клиринговой функции \eqref{E:Karmarkar} с эмпирическими данными \cite{Missbauer_2011}.}
\label{fig1}
\end{figure}

Примером эмпирических данных для построения КФ служат результаты наблюдения за работой принтера, предназначенного для нанесения графического изображения на отштампованные CD и DVD диски, приведенные в работе \cite{Missbauer_2011}. Выборочные данные получены от одной машины за сутки ee непрерывной работы и представляют собой точки в координатах <<WIP--выпуск>>. Координаты точек поделены нами на соответствующие максимальные величины, и нормированные таким образом эмпирические данные нанесены на Рис.~\ref{fig1} в виде кружков. Сплошной линией на графике показана теоретическая КФ \eqref{E:Karmarkar}, параметры $\widehat{\mathrm{TH}}$ и $K$ которой подобраны нелинейным методом наименьших квадратов из пакета \verb"lsqcurvefit" MATLAB.

КФ обычно выводится в предположении о стационарности процесса производства. Тем не менее, как исследователи, так и практики имеют склонность применить эту функцию к условиям, далеким от равновесия. Значительная часть современных работ по КФ признает необходимость четкой делимитации границ работоспособности КФ. В то же самое время --- насколько нам известно --- трудно указать хотя бы одну статью, где бы явно затрагивалась эта проблема. Заметное исключение представляет исследование Д.~Армбрустера \cite{Armbruster_2012b}, в котором на качественном уровне обсуждается применимость подхода КФ любого вида. В цитируемой работе утверждается, что КФ выражает собой квазистационарное (адиабатическое) приближение. Имеется в виду, что поток продукции на выходе мгновенно отслеживает малейшее колебание уровня WIP. Иными словами, КФ справедлива, если время релаксации флуктуаций числа работоспособных машин в производственной единице намного короче характерного времени изменения WIP. И все же, несмотря на всю конструктивность и ценность данного рассуждения, ему недостает ясно сформулированного количественного критерия.

Настоящая работа преследует двоякую цель: 1)~построить правдоподобную механистическую (то есть не регрессионную, а объясняющую механизм) модель <<элементарной>> производственной единицы в жидкостном приближении, из которой можно было бы вывести КФ; 2)~выявить условия справедливости полученной КФ в терминах эмпирически измеримых величин.

\bigskip
\textbf{2. Модель}

Основное предположение состоит в том, что трансформация материальных ресурсов в продукцию на промышленном предприятии происходит примерно по той же схеме, что и превращение молекул субстрата в молекулы других веществ в ходе ферментативного катализа. Действительно, в живой клетке молекула субстрата связывается с молекулой фермента, образуя короткоживущий промежуточный фермент-субстратный комплекс. Затем этот комплекс распадается на продукт и исходный фермент, способный катализировать новую реакцию (напр., \cite{Cornish_1979}).

На аналогию между трудом в производстве и катализатором в химической реакции впервые указал Н.~Джорджеску-Рёген \cite[p.~316--337]{Georgescu_1966}. В более общей форме, как отождествление основных фондов с ферментами, а вводимых факторов производства -- с субстратами, эта идея нашла свое отражение в работах И.\,А.~Полетаева \cite{Kolesova_1969} и Д.\,С.~Чернавского \cite[с.~134]{Romanovskii_1970}. Последующее приложение холистического подхода к моделированию процессов <<затраты-выпуск>> в сферах производства и услуг развивалось в публикациях \cite{Mustafin_1999, Milovanov_2001, Ponzi_2003, Niyirora_2017, MustKant_2019}.

Мы уподобляем ресурс субстрату, а конечное изделие --- продукту. В производственном цехе основные фонды (долгосрочные активы неизменного характера, предназначенные для выработки продуктов), такие как машинное оборудование, играют роль ферментов. С помощью машин и квалифицированной рабочей силы сырье, материалы, сборочные детали или полуфабрикаты совместно преобразуются в готовую продукцию.

Представим себе производственный цех, где происходит превращение одного ресурса в продукт. Это формально-математический объект, содержащий фиксированное количество одинаковых машин, обрабатывающих единицы ресурса, поступающие в систему, и накопитель (буфер), общий для всех машин, в котором находится ресурс в очереди на обработку.

Под единицей ресурса мы понимаем его количество, равное номинальной загрузке машины, указанной в ее технической документации для условий нормальной эксплуатации. Это позволяет исчислять ресурс и машины одинаковыми единицами измерения --- штуками.

В нашей модели такие производственные термины как <<материально-производствен\-ные запасы>> (inventory), <<материалы и заготовки, вовлеченные в технологический процесс производства>> (work in progress, WIP) и <<буферный запас>> (buffer stock) считаются синонимами, и все означают текущее количество ресурса в накопителе.

Машина может находится в одном из двух состояний: 1)~свободном для загрузки ресурсом или 2)~занятом обработкой ресурса. В каждый момент времени в машине на обработке может находиться только одна единица ресурса. Под обработкой понимается задержка ресурса на некоторое время в машине, пока идет трансформация ресурса в продукт. По окончании обработки комплекс <<машина--ресурс>> дезинтегрируется с высвобождением единицы продукта и возвращением машины в свободное состояние.

Поступление единиц ресурса в цех и длительность их обработки не зависят от того, сколько ресурса уже находится в цехе, или от каких-либо других факторов; продолжительность обработки ресурса не зависит от скорости его поступления в систему. Единица ресурса, поступившая на вход, может находиться в двух состояниях: 1)~обработки в машине или 2)~ожидания, если все имеющиеся машины заняты. Мы предполагаем, что на загрузку в машину единица ресурса выбирается из буфера случайным образом. Также предполагается, что емкость накопителя неограничена.

Перечисленные выше операции можно представить в виде псевдохимических уравнений
\begin{equation}\label{E:x-to-p}
\begin{gathered}
{\xymatrix@C-0.00pc@R-2.00pc{
\ar[r]^-{r}&{x} \ar[r]^(0.7){k_{1}}&{v} \ar[r]^(0.3){k_{2}} \ar@/^1.5pc/[d]&{p}\\
&&{u}\ar@/^1.5pc/[u]&
}}
\end{gathered}
\end{equation}
Здесь $x$ --- буферный запас ресурса в текущий момент времени, $u$ --- количество свободных машин, $v$ --- количество занятых машин, $p$ --- количество готового продукта. Как принято в химии, каждая стрелка указывает направление соответствующей стадии. Метками над стрелками обозначены кинетические константы стадий: $r$ --- скорость поступления ресурса (единица$\cdot$время$^{-1}$), $k_{1}$ --- скорость связывания единицы ресурса с машиной (единица$^{-1}\cdot$время$^{-1}$), $k_{2}$ --- число оборотов машины (время$^{-1}$), то есть максимальное количество единиц продукта, которое способна выпустить одна машина за единицу времени.

На схеме \eqref{E:x-to-p} закодированы как последовательность стадий, так и скорости, с которыми эти стадии протекают. Предполагается, что для потоков справедлив закон действующих масс, по которому скорость образования продукта с участием двух агентов пропорциональна произведению их количеств, иначе -- вероятности их встречи. В результате схеме \eqref{E:x-to-p} можно сопоставить систему уравнений материального баланса для всех участвующих агентов:
\begin{subequations}\label{E:basic}
\begin{alignat}{2}
\dot{x}&= -k_{1}x u +r,\quad& x(0)&=x_{0} \geqslant 0,\label{E:basic_x}\\
\dot{u}&= -k_{1}x u +k_{2}v,\quad& u(0)&=u_{0} > 0,\label{E:basic_u}\\
\dot{v}&= k_{1}x u -k_{2}v,\quad& v(0)&=0,\label{E:basic_v}\\
\dot{p}&= k_{2}v,\quad& p(0)&=0,\label{E:basic_y}
\end{alignat}
\end{subequations}
где точками сверху обозначены производные по времени $t$. Предполагается, что $r \geqslant 0$, $k_{1} > 0$ и $k_{2} > 0$.

\bigskip
\textbf{3. Анализ и результаты}
\paragraph{3.1. Случай $r>0$}
Складывая \eqref{E:basic_u} и \eqref{E:basic_v}, находим первый интеграл системы \eqref{E:basic}
\begin{equation}\label{E:u0}
u + v = u_{0} = \mathrm{const},
\end{equation}
смысл которого --- в сохранении общего количества машин.

Соотношение \eqref{E:u0} позволяет упростить систему \eqref{E:basic}, исключив из нее, скажем, $u$. Кроме того, поскольку уравнения \eqref{E:basic_x}, \eqref{E:basic_u} и \eqref{E:basic_v} не содержат $p$, то уравнение \eqref{E:basic_y} является подчиненным, и может рассматриваться отдельно после того, как получены решения для прочих переменных. В итоге остается система двух уравнений
\begin{equation}\label{E:basic_x-v}
\begin{aligned}
\dot{x}&= k_{1}x v -k_{1}u_{0}x +r,\\
\dot{v}&= -k_{1}x v +k_{1}u_{0}x -k_{2}v.
\end{aligned}
\end{equation}

Система \eqref{E:basic_x-v} имеет единственное стационарное состояние
\begin{equation}\label{E:ss}
(\bar{x},\bar{v}) = \biggl(\frac{K r}{k_{2} u_{0} -r},\frac{r}{k_{2}}\biggr),
\end{equation}
где $K=k_{2}/k_{1}$ --- новая константа, имеющая размерность ресурса. Заметим, что во избежание переполнения буфера на приток ресурса следует наложить ограничение $r < k_{2} u_{0}$. Стационарное состояние \eqref{E:ss} положительно и асимптотически устойчиво.

Определяя соответствующие характерные масштабы времени для $x$ и $v$
\begin{equation}\label{E:timescales}
T_{x}=(k_{1}\bar{v})^{-1},\quad T_{v}=(k_{1}\bar{x})^{-1},
\end{equation}
и переходя к новым безразмерным переменным и параметрам по формулам
\begin{gather}\label{E:param}
\xi = \frac{x}{\bar{x}},\quad \eta = \frac{v}{\bar{v}},\quad \tau = \frac{t}{T_{x}},\quad
a = \frac{u_{0}}{\bar{v}},\quad \varepsilon = \frac{T_{v}}{T_{x}} = \frac{\bar{v}}{\bar{x}},
\end{gather}
приводим систему \eqref{E:basic_x-v} к безразмерному виду
\begin{subequations}\label{E:xi-eta}
\begin{alignat}{2}
\dot{\xi} &= \xi\eta -a\xi +a -1\quad & \triangleq f(\xi,\eta),\label{E:xi-eta-1}\\
\varepsilon\,\dot{\eta} &= -\xi\eta +a\xi -(a -1)\eta\quad & \triangleq g(\xi,\eta),\label{E:xi-eta-2}
\end{alignat}
\end{subequations}
где точки сверху теперь означают дифференцирование по $\tau$, а $f$ и $g$ введены для краткого обозначения правых частей уравнений \eqref{E:xi-eta-1} и \eqref{E:xi-eta-2}.

Для типичного промышленного предприятия параметр $\varepsilon$ скорее всего мал. Так, например, в США оборудование и приспособления в серийном производстве используются в среднем лишь 25--35\% времени даже при полной мощности выпуска, а материалы, проходящие через металлообрабатывающий завод, проводят от 80 до 90\% времени в хранении и менее 5\% времени --- в обработке \cite{Bradt_1983}. Также известно \cite[p.~241]{Hopp_2008}, что на поточных линиях типичное соотношение <<WIP\,:\,машины>> приближается к 20:1. Так что можно с достаточной уверенностью полагать, что $\varepsilon \ll 1$, и в силу этого считать систему \eqref{E:xi-eta} сингулярно возмущенной.

Медленной переменной в \eqref{E:xi-eta} является $\xi$ (ресурс), быстрой --- $\eta$ (машины). Стандартной практикой упрощения сингулярно возмущенных (разнотемповых) систем служит применение метода медленного инвариантного многообразия (напр., \cite{Vasileva_1990, Verhulst_2005, Sobolev_2010}) и теоремы А.\,Н.~Тихонова \cite{Tikhonov_1952ru}.

Для автономной системы медленное инвариантное многообразие определяется как
\begin{equation*}
\eta(\varepsilon)=\phi(\xi,\varepsilon),
\end{equation*}
где функция $\phi$ --- разложение в ряд по степеням $\varepsilon$. Движение вдоль медленного многообразия дается медленным уравнением $\dot{\xi}=f(\xi,\phi(\xi,\varepsilon))$.

Технически для построения медленного инвариантного многообразия нужно найти члены асимптотического (внешнего) разложения
\begin{equation}\label{E:expansion}
\phi(\xi,\varepsilon) = \phi_{0}(\xi) +\varepsilon \phi_{1}(\xi) +\varepsilon^{2} \phi_{2}(\xi) +\ldots,
\end{equation}
где
\begin{equation*}
\phi_{0}(\xi) = \frac{a \xi}{a-1+\xi}
\end{equation*}
--- решение алгебраического уравнения $g(\xi,\eta)=0$ относительно $\eta$.

Подставляя \eqref{E:expansion} в \eqref{E:xi-eta-2} с учетом \eqref{E:xi-eta-1}, обозначая знаком штриха производную по $\xi$ и используя цепное правило дифференцирования, получаем
\begin{equation*}
\varepsilon\, \phi'(\xi,\varepsilon)\,f\bigl(\xi,\phi(\xi,\varepsilon)\bigr) = g\bigl(\xi,\phi(\xi,\varepsilon)\bigr),
\end{equation*}
или в развернутой форме
\begin{multline}
\varepsilon\,\biggl(\frac{(a-1)a}{(a-1+\xi)^2} +\varepsilon \phi_{1}^{\prime}(\xi) +\ldots\biggr)\\
\times\biggl[\xi\,\biggl(\frac{a \xi}{a-1+\xi} +\varepsilon \phi_{1}(\xi) +\ldots\biggr) +a(1-\xi)-1\biggr]\\
= -(a -1 +\xi)\biggl(\frac{a \xi}{a-1+\xi} +\varepsilon \phi_{1}(\xi) +\ldots\biggr) +a \xi.
\end{multline}
Собирая члены первого порядка по $\varepsilon$, находим выражение для $\phi_{1}$:
\begin{equation*}
\phi_{1}(\xi) = \frac{a(a-1)^3(\xi-1)}{(a-1+\xi)^4}.
\end{equation*}
Таким образом, с точностью до $\mathcal{O}(\varepsilon)$ искомое медленное инвариантное многообразие $\phi(\xi,\varepsilon)$ есть
\begin{equation}\label{slowman1}
\eta = \frac{a\xi}{a-1+\xi}\biggl(1 +\varepsilon\frac{(a-1)^3(\xi-1)}{\xi(a-1+\xi)^3}\biggr) +\mathcal{O}(\varepsilon^{2}).
\end{equation}
Многообразие является притягивающим, поскольку $\partial_{\eta}g = -(a-1+\xi) < 0$.

Поскольку $\eta = \phi_{0}(\xi)$ --- изолированный корень уравнения $g(\xi,\eta)=0$ и представляет собой устойчивую неподвижную точку уравнения \eqref{E:xi-eta-2}, и начальное условие $\eta(0)=0$ попадает в область его влияния, то в соответствии с теоремой Тихонова решение возмущенной системы \eqref{E:xi-eta} будет стремиться к решению уравнения $\dot{\xi}=f(\xi,\phi_{0}(\xi))$ при $\varepsilon \to 0$.

\begin{figure}[htb]
\noindent\centering{
\includegraphics[scale=1]{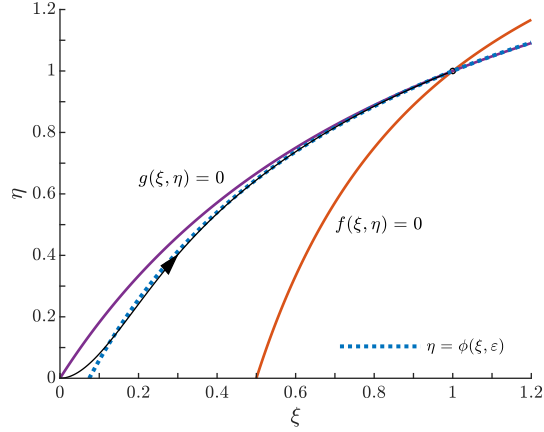}}
\caption{Фазовая плоскость системы \eqref{E:xi-eta}. Параметры: $a=2$, $\varepsilon = 0.1$.}
\label{fig2}
\end{figure}

На Рис.~\ref{fig2} можно видеть, как изображающая точка на фазовой плоскости $(\xi,\eta)$ относительно быстро <<садится>> на медленное инвариантное многообразие \eqref{slowman1} (в данном случае --- линию) и затем скользит по нему в направлении к устойчивому стационарному состоянию $(1,1)$. Переходный процесс до <<посадки>> имеет продолжительность порядка $\varepsilon$.

Ввиду заметного статистического разброса эмпирических наблюдений (см. Рис.~\ref{fig1}) на практике по-видимому вполне достаточно формул квазистационарного приближения нулевого порядка по $\varepsilon$. В размерном виде для мгновенного количества занятых машин можно написать
\begin{equation*}
v = \frac{u_{0} x}{K +x},
\end{equation*}
а для скорости потребления ресурса ---
\begin{equation*}
\dot{x} = r- \frac{k_{2}u_{0}x}{K +x}.
\end{equation*}
Соответственно скорость производства продукта будет иметь вид
\begin{equation}\label{E:M-M}
\dot{p} = \frac{k_{2}u_{0}x}{K +x}.
\end{equation}
Очевидно, что это выражение есть не что иное как КФ Кармаркара \eqref{E:Karmarkar} с точностью до обозначений. Формула \eqref{E:M-M} справедлива на масштабах времени порядка $K/r$ (см.~\eqref{E:timescales}).

\paragraph{3.2. Случай $r=0$}
Особая ситуация имеет место, когда внешнего притока ресурса нет, $r=0$, и в системе \eqref{E:basic_x-v} возможно только нулевое стационарное состояние $(0,0)$. Тем не менее формула \eqref{E:M-M} остается справедливой и в таком случае. Вводя новые переменные и параметры по формулам
\begin{gather*}
\xi = \frac{x}{x_{0}},\quad \eta = \frac{v}{u_{0}},\quad \tau = k_{1}u_{0} t,\quad b = \frac{K}{x_{0}},\quad \varepsilon = \frac{u_{0}}{K+x_{0}},
\end{gather*}
мы получаем следующие безразмерные уравнения:
\begin{equation}\label{E:nosupply}
\begin{alignedat}{2}
\dot{\xi} &= \xi\eta -\xi\quad & \triangleq f(\xi,\eta),\\
\varepsilon\,\dot{\eta} &= \frac{-\xi\eta +\xi -b\eta}{b+1}\quad  & \triangleq g(\xi,\eta).
\end{alignedat}
\end{equation}
При начальных условиях, удовлетворяющих неравенству $u_{0}/x_{0} \ll 1$, имеет место $\varepsilon \ll 1$, что делает систему \eqref{E:nosupply} сингулярно возмущенной.

Применяя к системе \eqref{E:nosupply} технику асимптотического разложения, использовавшуюся выше, находим в первом порядке по $\varepsilon$ медленное интегральное многообразие
\begin{equation}\label{slowman2}
\eta = \frac{\xi}{b+\xi}\biggl(1 +\varepsilon \frac{b^2(b+1)}{(b+\xi)^3}\biggr) +\mathcal{O}(\varepsilon^{2})
\end{equation}
и соответствующую скорость потребления ресурса
\begin{equation*}
\dot{\xi} = -\frac{b \xi}{b+\xi}\biggl(1 -\varepsilon \frac{b(b+1)\xi}{(b+\xi)^3}\biggr) +\mathcal{O}(\varepsilon^{2}).
\end{equation*}

\begin{figure}[htb]
\noindent\centering{
\includegraphics[scale=1]{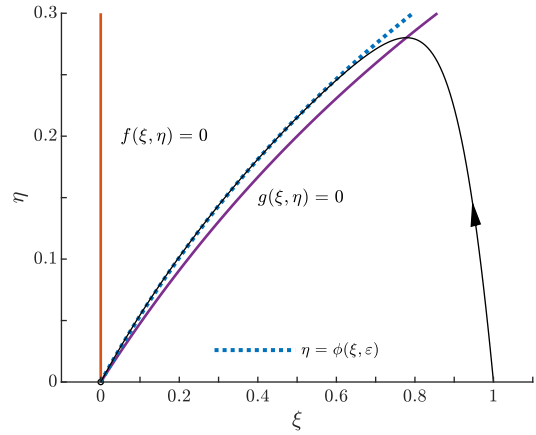}}
\caption{Фазовая плоскость системы \eqref{E:nosupply}. Параметры: $b=2$, $\varepsilon = 0.1$.}
\label{fig3}
\end{figure}

Рис.~\ref{fig3} иллюстрирует движение изображающей точки из начального положения $(1,0)$ сначала к медленному интегральному многообразию \eqref{slowman2} и потом вдоль него --- к равновесию в начале координат. В размерном виде нулевое приближение для КФ по-прежнему дается формулой \eqref{E:M-M}.

\bigskip
\textbf{4. Обсуждение и заключительные замечания}

Формула \eqref{E:M-M}, полученная в настоящей работе для КФ, идентична известному в ферментативной кинетике уравнению Михаэлиса--Ментен \cite{Johnson_2011}, ввиду подобия машины и фермента в кибернетическом смысле. В обоих процессах --- производстве изделия и биохимической реакции --- можно выделить три типа участников: объект на входе (ресурс, молекула субстрата), преобразователь (машина, фермент/катализатор) и трансформированный объект на выходе (готовое изделие, продукт реакции). При этом оба процесса идут через образование короткоживущего промежуточного комплекса, будь то машина--ресурс или фермент--субстрат.

Следует добавить, что формула \eqref{E:M-M} совпадает по форме также с известным в экологии функциональным откликом (functional response) хищника на жертву второго типа по К.~Холлингу \cite{Holling_1959}, $f(R) = aR/(1+a h R)$, где $f$ --- количество потребляемых жертв за единицу времени в расчете на одного хищника, $R$ --- плотность популяции жертв, $a$ --- частота успешных атак в расчете на одного хищника (время$^{-1}\cdot$хищник$^{-1}$), $h$ --- время, затрачиваемое хищником на разделку, поедание и усвоение одной жертвы (время$\cdot$хищник$\cdot$жертва$^{-1}$). Здесь роль трансформирующего агента играет хищник, преобразующий биомассу жертв в биомассу своей популяции, и точно так же малым параметром выступает соотношение численностей популяций хищников и жертв. При неограниченной доступности жертв частота их потребления одним хищником стремится к физиологическому максимуму: $f(+\infty) = 1/h$. В микробиологии отклик Холлинга второго типа обычно связывают с именем Ж.~Моно \cite{Monod_1949}, который ранее предложил такую же формулу для описания зависимости скорости потребления микроорганизмом субстрата от концентрации последнего.

Вывод КФ Кармаркара из <<химической>> модели \eqref{E:basic} отнюдь не ограничивается получением известного в исследовании операций результата новым способом. Общепринято считать, что КФ --- это инструмент для работы со стационарным режимом производства. Однако мы распространили понятие КФ также и на нестационарные режимы и указали количественные критерии ее применимости. Фактически это условия, при которых справедливо квазистационарное приближение, или когда оправданно использование медленного интегрального многообразия. Как показано, малость характерного соотношения <<машины\,:\,WIP>> практически гарантирует применимость приближения КФ. Малость указанного параметра порождает наличие двух сильно различающихся характерных масштабов времени в производственной единице: более длинного, соответствующего динамике WIP (наличного ресурса), и более короткого, соответствующего динамике машин, вовлеченных в производство. Существование такой иерархии временн\'{ы}х шкал делает возможным квазистационарное поведение популяции занятых машин путем быстрой подстройки к текущему объему ожидающего ресурса.

Модель \eqref{E:basic} в равной степени пригодна также для описания процесса обслуживания, если считать, что $r$ --- поток клиентов (заявок), $x$ --- спрос (очередь) на услугу, а $u$ и $v$ --- соответственно свободные и занятые обслуживающие агенты (например, продавцы или операторы колл-центра). В теории очередей известно \cite[p.~84]{Whitt_2018}, что время ожидания может часто превышать время обслуживания, поэтому квазистационарное приближение должно работать и в такой системе \cite{MustKant_2019}.

Наше последнее замечание касается возможной стохастичности притока ресурса. В проведенном анализе предполагалось, что $r = \mathrm{const}$. Рассмотрение соответствующего уравнения Ланжевена со случайно флуктуирующим аддитивным членом выходит за рамки настоящего исследования. Для краткости ограничимся лишь комментарием, что клиринговая функция сохранит работоспособность и в этой ситуации, если время корреляции флуктуаций притока ресурса (например, сезонных вариаций) окажется намного длиннее характерного времени релаксации популяции машин.

\bigskip
\textbf{Литература}
\providecommand*{\BibDash}{}
\renewcommand\refname{\vskip -1cm}

\vspace{5mm}
\noindent
\footnotesize
\textbf{Сведения об авторах:}\\
\textit{Алмаз Тлемисович Мустафин} --- д-р техн. наук, проф.; \verb"a.mustafin@satbayev.university"\\
\textit{Алия Кажбековна Кантарбаева} --- д-р экон. наук, проф.; \verb"kantarbayeva.aliya@kaznu.kz"

\normalsize
\bigskip
\noindent
\large{Clearing function in the context of the invariant manifold method}\\

\noindent
\textit{A. Mustafin}$^1$, \textit{A. Kantarbayeva}$^2$\\
\small{$^1$Satbayev University, 22~Satbayev~St., Almaty, 050013, Kazakhstan}\\
\small{$^2$al-Farabi Kazakh National University, 71~al-Farabi~Ave., Almaty, 050040, Kazakhstan}
\vspace{2mm}

Clearing functions (CFs), which express a mathematical relationship between the expected throughput of a production facility in a planning period and its workload (or work-in-progress, WIP) in that period have shown considerable promise for modeling WIP-dependent cycle times in production planning. While steady-state queueing models are commonly used to derive analytic expressions for CFs, the finite length of planning periods calls their validity into question. We apply a different approach to propose a mechanistic model for one-resource, one-product factory shop based on the analogy between the operation of machine and enzyme molecule. The model is reduced to a singularly perturbed system of two differential equations for slow (WIP) and fast (busy machines) variables, respectively. The analysis of this slow-fast system finds that CF is nothing but a result of the asymptotic expansion of the slow invariant manifold. The validity of CF is ultimately determined by how small is the parameter multiplying the derivative of the fast variable. It is shown that sufficiently small characteristic ratio 'working machines\,:\,WIP' guarantees the applicability of CF approximation in unsteady-state operation.

\textit{Keywords:} work in progress, production model, quasi-steady-state approximation, singular perturbation, enzyme catalysis

\bigskip


\vspace{5mm}
\noindent
\footnotesize
\textbf{Authors' information:}\\
\textit{Almaz Mustafin} --- DSc in Engineering, Professor; \verb"a.mustafin@satbayev.university"\\
\textit{Aliya Kantarbayeva} --- DSc in Economics, Professor; \verb"kantarbayeva.aliya@kaznu.kz"

\end{document}